\documentclass[aps, prb, reprint, superscriptaddress]{revtex4-1}
\usepackage[english]{babel}
\usepackage{amsmath,amsthm}
\usepackage{amsfonts}
\usepackage{xspace}
\usepackage{bm}
\usepackage{booktabs}
\usepackage{graphicx}
\usepackage{dcolumn}% Align table columns on decimal point
\usepackage[mathlines]{lineno}% Enable numbering of text and display math
\usepackage[colorlinks]{hyperref}  
\usepackage{gensymb}
\usepackage{epstopdf}
\usepackage{newfloat}
\usepackage{color,soul}
\DeclareFloatingEnvironment[name={Figure S}]{suppfigure}
\DeclareFloatingEnvironment[name={S}]{suppEquation}

\newcommand*{\citen}[1]{%
  \begingroup
    \romannumeral-`\x % remove space at the beginning of \setcitestyle
    \setcitestyle{numbers}%
    \cite{#1}%
  \endgroup   
}

\begin{document}

\title{Comparison of spin-orbit torques and spin pumping across NiFe/Pt and NiFe/Cu/Pt interfaces}%

\author{Tianxiang Nan}%
\affiliation{ 
Department of Electrical and Computer Engineering, Northeastern University, Boston, MA 02115%\\This line break forced with \textbackslash\textbackslash
}%
\author{Satoru Emori}%
\email{
s.emori@neu.edu
}
\affiliation{ 
Department of Electrical and Computer Engineering, Northeastern University, Boston, MA 02115%\\This line break forced with \textbackslash\textbackslash
}%
\author{Carl T. Boone}%
\affiliation{ 
Department of Physics, Boston University, Boston, MA 02215%\\This line break forced with \textbackslash\textbackslash
}%
\author{Xinjun Wang}%
\affiliation{ 
Department of Electrical and Computer Engineering, Northeastern University, Boston, MA 02115%\\This line break forced with \textbackslash\textbackslash
}%
\author{Trevor M. Oxholm}%
\affiliation{ 
Department of Electrical and Computer Engineering, Northeastern University, Boston, MA 02115%\\This line break forced with \textbackslash\textbackslash
}%
\author{John G. Jones}%
\affiliation{ 
Materials and Manufacturing Directorate, Air Force Research Laboratory, Wright-Patterson AFB, OH 45433%\\This line break forced with \textbackslash\textbackslash
}%
\author{Brandon M. Howe}%
\affiliation{ 
Materials and Manufacturing Directorate, Air Force Research Laboratory, Wright-Patterson AFB, OH 45433%\\This line break forced with \textbackslash\textbackslash
}%
\author{Gail J. Brown}%
\affiliation{ 
Materials and Manufacturing Directorate, Air Force Research Laboratory, Wright-Patterson AFB, OH 45433%\\This line break forced with \textbackslash\textbackslash
}%
\author{Nian X. Sun}%
\email{
n.sun@neu.edu
}
\affiliation{ 
Department of Electrical and Computer Engineering, Northeastern University, Boston, MA 02115%\\This line break forced with \textbackslash\textbackslash
}%
\date{May 25, 2015}% It is always \today, today,
             %  but any date may be explicitly specified

% ----------------------------------------------------------------
\begin{abstract}
We experimentally investigate spin-orbit torques and spin pumping in NiFe/Pt bilayers with direct and interrupted interfaces.
The damping-like and field-like torques are simultaneously measured with spin-torque ferromagnetic resonance tuned by a dc bias current, whereas spin pumping is measured electrically through the inverse spin Hall effect using a microwave cavity.  
Insertion of an atomically thin Cu dusting layer at the interface reduces the damping-like torque, field-like torque, and spin pumping by nearly the same factor of $\approx$1.4.
This finding confirms that the observed spin-orbit torques predominantly arise from diffusive transport of spin current generated by the spin Hall effect.
We also find that spin-current scattering at the NiFe/Pt interface contributes to additional enhancement in magnetization damping that is distinct from spin pumping.  
\end{abstract}
\maketitle
% ----------------------------------------------------------------
\section{Introduction}\label{sec:intro}
Current-induced torques due to spin-orbit effects~\cite{Brataas2014, Gambardella2011a, Haney2013a} potentially allow for more efficient control of magnetization than the conventional spin-transfer torques~\cite{Ralph2008, Brataas2012a}. 
The spin Hall effect~\cite{Hoffmann2013} is reported to be the dominant source of spin-orbit torques in thin-film bilayers consisting of a ferromagnet (FM) interfaced with a normal metal (NM) with strong spin-orbit coupling.  
Of particular technological interest is the spin-Hall ``damping-like'' torque that induces magnetization switching~\cite{Miron2011, Liu2012, Bhowmik2014, Yu2014b}, domain-wall motion~\cite{Haazen2013, Emori2014d, Ryu2014, Ueda2014a}, and high-frequency magnetization dynamics~\cite{Demidov2012, Liu2012a, Liu2013, Duan2014a, Ranjbar2014, Hamadeh2014a}. 
While this spin-Hall torque originates from spin-current generation within the bulk of the NM layer, the magnitude of the torque depends on the transmission of spin current across the FM/NM interface~\cite{Haney2013a}.  
Some FM/NM bilayers with $\sim$1-nm thick FM exhibit another spin-orbit torque that is phenomenologically identical to a torque from an external magnetic field~\cite{Kim2013a, Garello2013, Fan2013, Fan2014, Pai2014, Pai2014a, Emori2014c, Woo2014}.  
This ``field-like'' torque is also interface-dependent, because it may emerge from the Rashba effect at the FM/NM interface~\cite{Gambardella2011a}, or the nonadiabaticity~\cite{Ralph2008} of spin-Hall-generated spin current transmitted across the interface~\cite{Haney2013a, Fan2013, Fan2014, Pai2014}.  

To understand the influence of the FM/NM interface on magnetization dynamics, many studies have experimentally investigated resonance-driven spin pumping from FM to NM~\cite{Tserkovnyak2002a, Tserkovnyak2002}, detected with enhanced damping~\cite{Ghosh2011, Boone2013, Boone2014, Heinrich2011, Sun2013a} or dc voltage due to the inverse spin Hall effect~\cite{Azevedo2005a, Saitoh2006, Mosendz2010a, Czeschka2011a, Ando2011, Deorani2013, Weiler2014c, Obstbaum2014, Rojas-Sanchez2014, Wang2014}. 
The parameter governing spin-current transmission across the FM/NM interface is the spin-mixing conductance $G_{\uparrow\downarrow}$ (Ref.~\citen{Brataas2000}).
%~\footnote{The spin-mixing ``conductivity'' $G_{\uparrow\downarrow}$ is related to the oft-used spin-mixing ``conductance'' $g_{\uparrow\downarrow}$ by $g_{\uparrow\downarrow}=(h/e^2)G_{\uparrow\downarrow}$}. 
Simultaneously investigating spin pumping and spin-orbit torques, which are theoretically reciprocal effects~\cite{Brataas2012a}, should reveal the interface dependence of the observed torques in FM/NM.  

Here we investigate spin-orbit torques and magnetic resonance in in-plane magnetized NiFe/Pt bilayers with direct and interrupted interfaces.  
To modify the NiFe/Pt interface, we insert an atomically thin dusting layer of Cu that does not exhibit strong spin-orbit effects by itself.  
We use spin-torque ferromagnetic resonance (ST-FMR)~\cite{Sankey2007,Liu2011} combined with dc bias current to extract the damping-like and field-like torques simultaneously. 
We also independently measure the dc voltage generated by spin pumping across the FM/NM interface.  
The interfacial dusting reduces the damping-like torque, field-like torque, and spin pumping by the same factor.
This finding is consistent with the diffusive spin-Hall mechanism~\cite{Haney2013a, Boone2013} of spin-orbit torques, where spin transfer between NM and FM depends on the interfacial spin-mixing conductance.  

\begin{figure*}
\includegraphics[width=1.1\columnwidth]{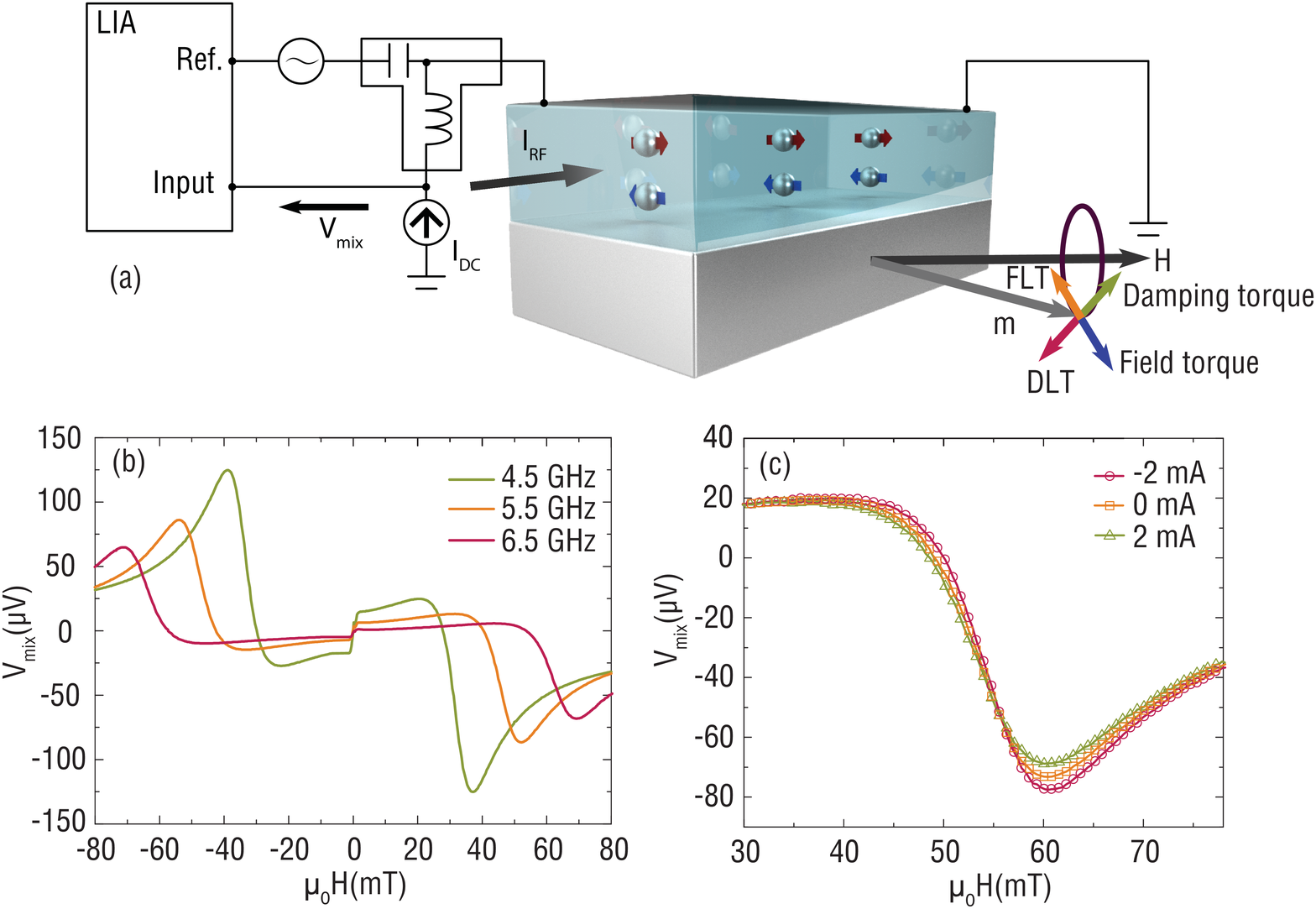}
\caption{\label{fig:STFMR}(a) Schematic of the dc-tuned spin-torque ferromagnetic resonance (ST-FMR) setup and the symmetry of torques acting on the magnetization $\mathbf{m}$. Through spin-orbit effects, the charge current in the normal metal generates two torques: damping-like torque (DLT) and field-like torque (FLT). (b,c) ST-FMR spectra of NiFe/Pt at different frequencies (b) and dc bias currents (c).} 
\end{figure*}

\section{Experimental Details}\label{sec:exp}
\subsection{Samples}\label{subsec:samples}
The two film stacks compared in this study are \textit{sub}/Ta(3)/Ni$_{80}$Fe$_{20}$(2.5)/Pt(4) (``NiFe/Pt'') and \textit{sub}/Ta(3)/Ni$_{80}$Fe$_{20}$(2.5)/Cu(0.5)/Pt(4)~(``NiFe/Cu/Pt''),  where the numbers in parentheses are nominal layer thicknesses in nm and \textit{sub} is a Si(001) substrate with a 50-nm thick SiO$_2$ overlayer.  
All layers were sputter-deposited at an Ar pressure of $3\times10^{-3}$ Torr with a background pressure of $\lesssim$1$\times10^{-7}$ Torr.
The atomically thin dusting layer of Cu modifies the NiFe/Pt interface with minimal current shunting.  
The Ta seed layer facilitates the growth of thin NiFe with narrow resonance linewidth and near-bulk saturation magnetization~\cite{Ghosh2011, Boone2014}. 

We measured the saturation magnetization $M_s=(5.8\pm0.4)\times10^5$ A/m for both NiFe/Pt and NiFe/Cu/Pt with vibrating sample magnetometry.
From four-point measurements on various film stacks and assuming that individual constituent layers are parallel resistors, we estimate the resistivities of Ta(3), NiFe(2.5), Cu(0.5), and Pt(4) to be 240 $\mu\Omega$cm, 90 $\mu\Omega$cm, 60 $\mu\Omega$cm, and 40 $\mu\Omega$cm, respectively.  
Approximately 70\% of the charge current thus flows in the Pt layer.
In the subsequent analysis, we also include the small damping-like torque and the Oersted field from the highly resistive Ta layer (see Appendix A). 

\subsection{Spin-torque ferromagnetic resonance}\label{subsec:STFMR}
We fabricated 5-$\mu$m wide, 25-$\mu$m long microstrips of NiFe/Pt and NiFe/Cu/Pt with Cr/Au ground-signal-ground electrodes using photolithography and liftoff.  
We probed magnetization dynamics in the microstrips using ST-FMR (Refs.~\citen{Sankey2007, Liu2011}) as illustrated in Fig.~\ref{fig:STFMR}(a): an rf current drives resonant precession of magnetization in the bilayer, and the rectified anisotropic magnetoresistance voltage generates an FMR spectrum.  
The rf current power output was +8 dBm and modulated with a frequency of 437 Hz to detect the rectified voltage using a lock-in amplifier. 
The ST-FMR spectrum (e.g., Fig.~\ref{fig:STFMR}(b)) was acquired at a fixed rf driving frequency by sweeping an in-plane magnetic field $|\mu_0 H|<80$ mT applied at an angle $|\phi| = 45\degree$ from the current axis.  
The rectified voltage $V_{mix}$ constituting the ST-FMR spectrum is fit to a Lorentzian curve of the form
\begin{equation}\label{eq:lorentzian}
\begin{split}
V_{mix} = & S\frac{W^2}{(\mu_0H-\mu_0H_{FMR})^2+W^2}\\ &+A\frac{W(\mu_0H-\mu_0H_{FMR})}{(\mu_0H-\mu_0H_{FMR})^2+W^2},
\end{split}
\end{equation}
where $W$ is the half-width-at-half-maximum resonance linewidth, $H_{FMR}$ is the resonance field,   $S$ is the symmetric Lorentzian coefficient, and $A$ is the antisymmetric Lorentzian coefficient.
Representative fits are shown in Fig.~\ref{fig:STFMR}(c). 

The lineshape of the ST-FMR spectrum, parameterized by the ratio of $S$ to $A$ in Eq.~\ref{eq:lorentzian}, has been used to evaluate the ratio of the damping-like torque to the net effective field from the Oersted field and field-like torque~\cite{Liu2011, Kondou2012a, Skinner2014, Wang2014c, Mellnik2014, Pai2014a}.
To decouple the damping-like torque from the field-like torque, the magnitude of the rf current in the bilayer would need to be known~\cite{Liu2011, Wang2014c}.  
Other contributions to $V_{mix}$ (Refs.~\citen{Yamaguchi2007a, Ganguly2014, Kasai2014}) may also affect the analysis based on the ST-FMR lineshape.  

\begin{figure*}
\includegraphics[width=1.25\columnwidth]{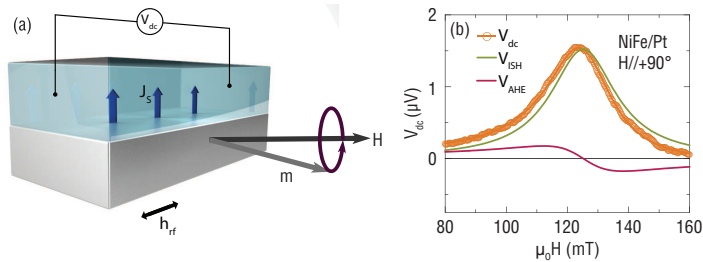}
\caption{\label{fig:ISHcartoon}(a) Schematic of the dc spin-pumping (inverse spin Hall effect) voltage measurement. (b) Representative dc voltage spectrum. The inverse spin Hall signal $V_{ISH}$ dominates the anomalous Hall effect signal $V_{AHE}$.} 
\end{figure*}

We use a modified approach where an additional dc bias current $I_{dc}$ in the bilayer, illustrated in Fig.~\ref{fig:STFMR}(a), transforms the ST-FMR spectrum as shown in Fig.~\ref{fig:STFMR}(c).  
A high-impedance current source outputs $I_{dc}$, and we restrict $|I_{dc}|\leq2$ mA (equivalent to the current density in Pt $|J_{c,Pt}|~<~10^{11}$ A/m$^2$) to minimize Joule heating and nonlinear dynamics.   
The dependence of the resonance linewidth $W$ on $I_{dc}$ allows for quantification of the damping-like torque~\cite{Ando2008a, Liu2011, Demidov2011, Pai2012, Ganguly2014, Kasai2014, Duan2014b, Emori2015}, while the change in the resonance field $H_{FMR}$ yields a direct measure of the field-like torque~\cite{Mellnik2014}. 
Thus, dc-tuned ST-FMR quantifies both spin-orbit torque contributions.

\subsection{Electrical detection of spin pumping}\label{subsec:spinpumping}
\begin{figure*}
\includegraphics[width=1.8\columnwidth]{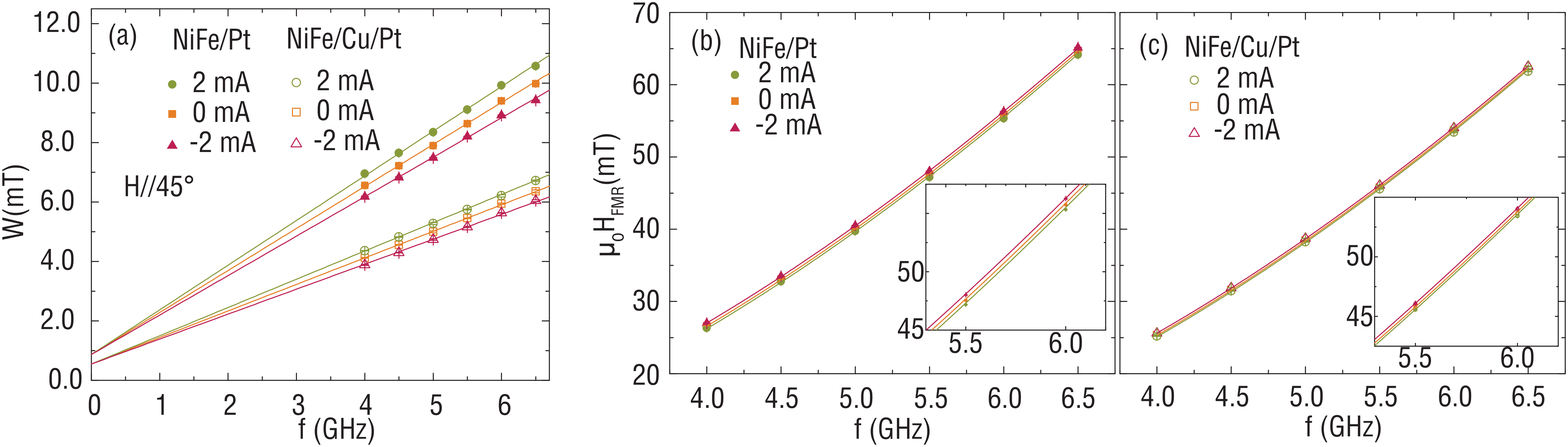}
\caption{\label{fig:broadband}(a) Resonance linewidth $W$ versus frequency $f$ at different dc bias currents. (b,c) Resonance field $H_{FMR}$ versus frequency $f$ at different dc-bias currents for NiFe/Pt (b) and NiFe/Cu/Pt (c).} 
\end{figure*}
The inverse spin Hall voltage $V_{ISH}$ due to spin pumping was measured in 100-$\mu$m wide, 1500-$\mu$m long strips of NiFe/Pt and NiFe/Cu/Pt with Cr/Au electrodes attached on both ends, similar to the sub-mm wide strips used in Ref.~\citen{Emori2015}.
These NiFe/(Cu/)Pt strips were fabricated on the same substrate as the ST-FMR device sets described in Sec.~\ref{subsec:STFMR}.  
The sample was placed in the center of a rectangular TE$_{102}$ microwave cavity operated at a fixed rf excitation frequency of 9.55 GHz and rf power of 100 mW.  
A bias field $H$ was applied within the film plane and transverse to the long axis of the strip. 
The dc voltage $V_{dc}$ across the sample was measured using a nanovoltmeter while sweeping the field, as illustrated in Fig.~\ref{fig:ISHcartoon}(a).
The acquired $V_{dc}$ spectrum is fit to Eq.~\ref{eq:lorentzian} as shown by a representative result in Fig.~\ref{fig:ISHcartoon}(b).
The inverse spin Hall voltage is defined as the amplitude of the symmetric Lorentzian coefficient $S$ in Eq.~\ref{eq:lorentzian} (Refs.~\citen{Mosendz2010a, Czeschka2011a, Ando2011, Deorani2013, Rojas-Sanchez2014}).
We note that the antisymmetric Lorentzian coefficient is substantially smaller, indicating that the voltage signal from the inverse spin Hall effect dominates over that from the anomalous Hall effect. 

\begin{figure*}
\includegraphics[width=1.4\columnwidth]{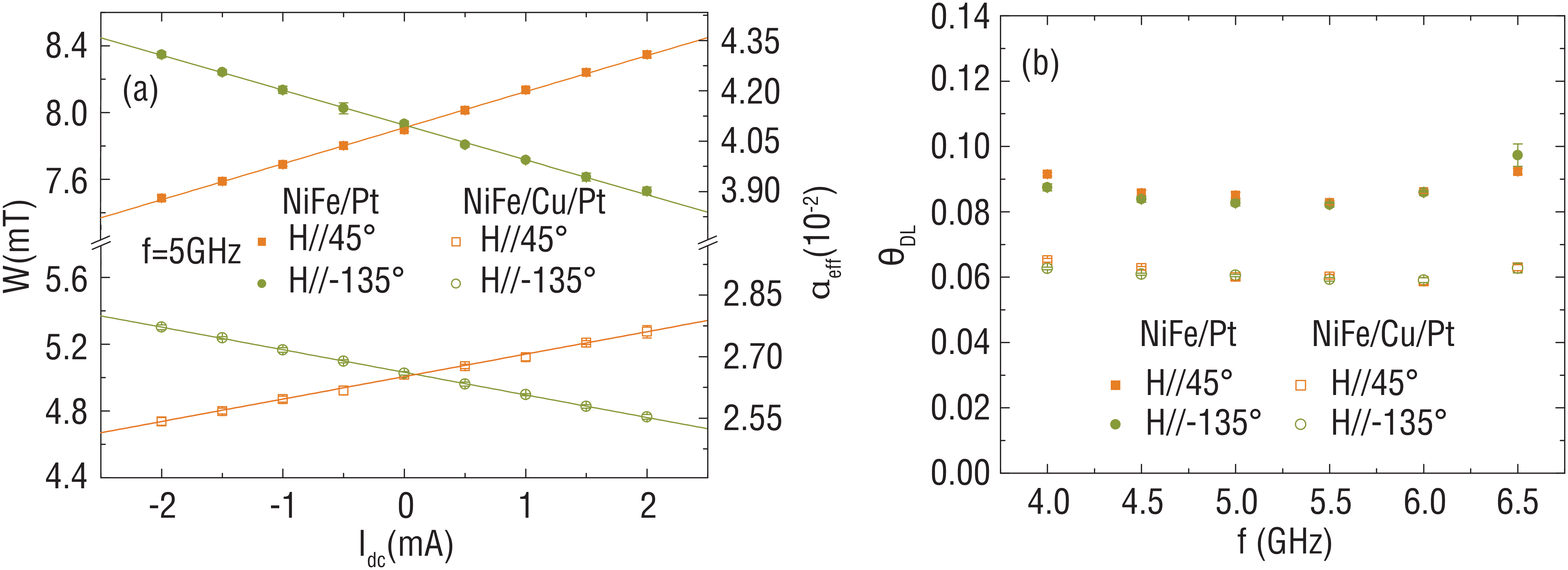}
\caption{\label{fig:dampinglike}(a) Resonance linewidth $W$ versus dc bias current $I_{dc}$ at $f=5$ GHz. (b) Effective spin Hall angle $\theta_{DL}$ calculated at several frequencies.} 
\end{figure*}

\section{Results and Analysis}
\subsection{Magnetic resonance properties}
Fig.~\ref{fig:broadband}(a) shows the plot of the ST-FMR linewidth $W$ as a function of frequency $f$ for NiFe/Pt and NiFe/Cu/Pt at $I_{dc} = 0$ and $\pm2$ mA.  
The Gilbert damping parameter $\alpha$ is calculated for each sample in Fig.~\ref{fig:broadband}(a) from 
\begin{equation}\label{eq:damping}
W = W_0 + \frac{2\pi\alpha}{|\gamma|}f,
\end{equation}
where $W_0$ is the inhomogeneous linewidth broadening, $f$ is the frequency, and $\gamma$ is the gyromagnetic ratio. 
With the Land\'{e} g-factor $g_L = 2.10$ for NiFe (Refs.~\citen{Mizukami2002, Ghosh2011, Weiler2014c, Boone2014}), $|\gamma|/2\pi = $(28.0 GHz/T)$\cdot (g_L/2) = 29.4$ GHz/T.  
From the slope in Fig.~\ref{fig:broadband}(a) at $I_{dc} = 0$, $\alpha = 0.043\pm0.001$ for NiFe/Pt and $\alpha = 0.027\pm0.001$ for NiFe/Cu/Pt.   
The reduction in damping with interfacial Cu-dusting is consistent with prior studies on FM/Pt with nm-thick Cu insertion layers~\cite{Weiler2014c, Ghosh2011, Boone2014, Rojas-Sanchez2014, Sun2013a}.  

A fit of $H_{FMR}$ versus frequency at $I_{dc} = 0$ to the Kittel equation
\begin{equation}\label{eq:kittel}
\begin{split}
\mu_0 H_{FMR} = 
&\tfrac{1}{2}\left(-\mu_0 M_{eff} +\sqrt{(\mu_0 M_{eff})^2+4(f/\gamma)^2}\right) \\
&-\mu_0 H_k+\mu_0 \Delta H_{FMR}(I_{dc}),
\end{split}
\end{equation}
shown in Figs.~\ref{fig:broadband}(b),(c), gives the effective magnetization $M_{eff} = 5.6\times10^5$ A/m for NiFe/Pt and $5.9\times10^5$ A/m for NiFe/Cu/Pt, with the in-plane anisotropy field $|\mu_0 H_k|<1$~mT.
$M_{eff}$ and $M_{s}$ are indistinguishable within experimental uncertainty, implying negligible perpendicular magnetic anisotropy in NiFe/(Cu/)Pt. 

When $I_{dc} \neq 0$, the linewidth $W$ is reduced for one current polarity and enhanced for the opposite polarity, as shown in  Fig.~\ref{fig:broadband}(a).  
The empirical damping parameter defined by Eq.~\ref{eq:damping} changes with $I_{dc}$ (see Appendix B), which indicates the presence of a current-induced damping-like torque.  
Similarly, $I_{dc} \neq 0$ generates an Oersted field and a spin-orbit field-like torque that together shift the resonance field $H_{FMR}$ as shown in Figs.~\ref{fig:broadband}(b),(c).
We discuss the quantification of the damping-like torque in Sec.~\ref{subsec:DLT} and the field-like torque in Sec.~\ref{subsec:FLT}.

\subsection{Damping-like torque}\label{subsec:DLT}
Fig.~\ref{fig:dampinglike}(a) shows the linear change in $W$ as a function of $I_{dc}$ at a fixed rf frequency of 5 GHz.  
Reversing the external field (from $\phi =45\degree$ to -135\degree) magnetizes the sample in the opposite direction and reverses the polarity of the damping-like torque. 

$W$ is related to the current-dependent effective damping parameter $\alpha_{eff}$ at fixed $f$, $\alpha_{eff} = |\gamma|/(2\pi f) (W~-~W_0)$. 
The magnitude of the damping-like torque is parameterized by the effective spin Hall angle $\theta_{DL}$, proportional to the ratio of the spin current density $J_s$ crossing the FM/NM interface to the charge current density $J_c$ in Pt.
$\theta_{DL}$ at each frequency, plotted in Fig.~\ref{fig:dampinglike}(b), is calculated from the $I_{dc}$ dependence of $\alpha_{eff}$ (Refs.~\citen{Petit2007, Liu2011}):
\begin{equation}\label{eq:JsJc}
|\theta_{DL}| = \frac{2|e|}{\hbar} \frac{\left( H_{FMR}+\tfrac{M_{eff}}{2} \right) \mu_0 M_s t_{F}}{|\sin\phi|} \left| \frac{\Delta\alpha_{eff}}{\Delta J_{c}} \right|,
\end{equation}
where $t_F$ is the FM thickness. 
Assuming that the effective spin Hall angle is independent of frequency, we find $\theta_{DL} = 0.087\pm0.007$ for NiFe/Pt and $\theta_{DL} = 0.062\pm0.005$ for NiFe/Cu/Pt. 
These values are similar to recently reported $\theta_{DL}$ in NiFe/Pt bilayers~\cite{Czeschka2011a, Weiler2014c, Liu2011, Wang2014c, Ganguly2014, Kasai2014, Ando2008a, Duan2014b}.  

\begin{figure*}
\includegraphics[width=1.15\columnwidth]{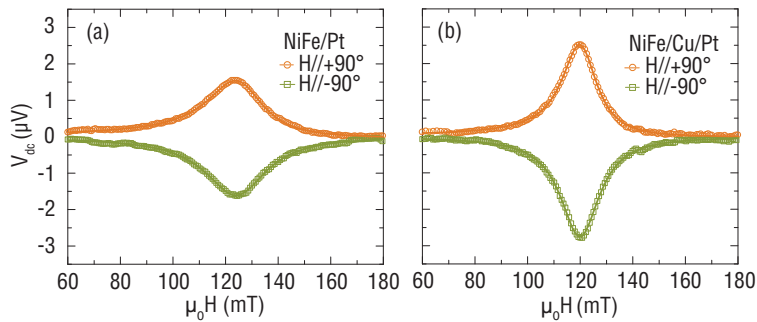}
\caption{\label{fig:ISH}(a,b) dc voltage $V_{dc}$ spectra, dominated by the inverse spin Hall voltage $V_{SH}$, measured around resonance in NiFe/Pt (a) and NiFe/Cu/Pt (b).} 
\end{figure*}

$\theta_{DL}$ of NiFe/(Cu/)Pt is related to the intrinsic spin Hall angle $\theta_{SH}$ of Pt through the spin diffusion theory used in Refs.~\citen{Boone2013, Haney2013a}.
For a Pt layer much thicker than its spin diffusion length $\lambda_{Pt}$, $\theta_{DL}$ is proportional to the real part of the effective spin-mixing conductance $G_{\uparrow\downarrow}^{eff}$,
\begin{equation}\label{eq:GSHE}
\theta_{DL} = \frac{2\text{Re}[G_{\uparrow\downarrow}^{eff}]}{{\sigma_{Pt}}/{\lambda_{Pt}}}\theta_{SH} ,
\end{equation}
where $\sigma_{Pt}$ is the conductivity of the Pt layer and $G_{\uparrow\downarrow}^{eff}~=~ G_{\uparrow\downarrow}(\sigma_{Pt}/\lambda_{Pt})/(2G_{\uparrow\downarrow}+\sigma_{Pt}/\lambda_{Pt})$ includes the spin-current backflow factor~\cite{Tserkovnyak2002, Boone2013}.  
Assuming that $\lambda_{Pt}$, $\sigma_{Pt}$, and $\theta_{SH}$ in Eq.~\ref{eq:GSHE} are independent of the interfacial Cu dusting layer, $G_{\uparrow\downarrow}^{eff}$ is a factor of $1.4\pm0.2$ greater for NiFe/Pt than NiFe/Cu/Pt based on the values of $\theta_{DL}$ found above.  

\subsection{Reciprocity of damping-like torque and spin pumping}\label{subsec:ISH}
Fig.~\ref{fig:ISH} shows representative results of the dc inverse spin Hall voltage induced by spin pumping, each fitted to the Loretzian curve defined by Eq.~\ref{eq:lorentzian}.   
Reversing the bias field reverses the moment orientation of the pumped spin current and thus inverts the polarity of $V_{ISH}$, consistent with the mechanism of the inverse spin Hall effect. 
By averaging measurements at opposite bias field polarities for different samples, we find $|V_{ISH}|=1.5\pm0.2$ $\mu$V for NiFe/Pt and $|V_{ISH}|=2.6\pm0.2$ $\mu$V for NiFe/Cu/Pt.  

The inverse spin Hall voltage $V_{ISH}$ is given by~\cite{Mosendz2010a} 
\begin{equation}\label{eq:VISH}
|V_{ISH}| = 
\frac{h}{|e|}
G_{\uparrow\downarrow}^{eff}|\theta_{SH}|\lambda_{Pt}
\tanh\left(\frac{t_{Pt}}{2\lambda_{Pt}}\right)fR_sLP
 \left(\frac{\gamma h_{rf}}{2\alpha\omega}\right)^2,
\end{equation}
where $R_s$ is the sheet resistance of the sample, $L$ is the length of the sample, $P$ is the ellipticity parameter of magnetization precession, and $h_{rf}$ is the amplitude of the microwave excitation field.  
The factor ${\gamma h_{rf}}/{2\alpha\omega}$ is equal to the precession cone angle at resonance in the linear (small angle) regime.  
By collecting all the factors in Eq.~\ref{eq:VISH} that are identical for NiFe/Pt and NiFe/Cu/Pt into a single coefficient $C_{ISH}$, Eq.~\ref{eq:VISH} is rewritten as
\begin{equation}\label{eq:VISHsimp}
|V_{ISH}| = C_{ISH}
\frac{R_sG_{\uparrow\downarrow}^{eff}}{\alpha^2}.
\end{equation}
We note that the small difference in $M_{eff}$ for NiFe/Pt and NiFe/Cu/Pt yields a difference in $P$ (Eq.~\ref{eq:VISH}) of $\sim$1\%, which we neglect here.  

From Eq.~\ref{eq:VISHsimp}, we estimate that $G_{\uparrow\downarrow}^{eff}$ of the NiFe/Pt interface is greater than that of the NiFe/Cu/Pt interface by a factor of $1.4\pm0.2$.  
The dc-tuned ST-FMR and dc spin-pumping voltage measurements therefore yield quantitatively consistent results, confirming the reciprocity between the damping-like torque (driven by the direct spin Hall effect) and spin pumping (detected with the inverse spin Hall effect).  
The fact that the diffusive model captures the observations supports the spin-Hall mechanism leading to the damping-like torque. 

\subsection{Interfacial damping and spin-current transmission}
Provided that the enhanced damping $\alpha$ in NiFe/(Cu/)Pt (Fig.~\ref{fig:broadband}(a)) is entirely due to spin pumping into the Pt layer, the real part of the interfacial spin-mixing conductance can be calculated by 
\begin{equation}\label{eq:Geff}
\text{Re}[G_{\uparrow\downarrow}^{eff}] = \frac{2e^2 M_s t_F}{\hbar^2 |\gamma|}(\alpha-\alpha_0).  
\end{equation}
Using $\alpha_0 = 0.011$ measured for a reference film stack \textit{sub}/Ta(3)/NiFe(2.5)/Cu(2.5)/TaOx(1.5) with negligible spin pumping into the top NM layer of Cu, we obtain Re$[G_{\uparrow\downarrow}^{eff}] = (11.6\pm0.9)\times10^{14}$~$\Omega^{-1}$m$^{-2}$ for NiFe/Pt and $(5.8\pm0.5)\times10^{14}$~$\Omega^{-1}$m$^{-2}$ for NiFe/Cu/Pt.  
This factor of 2 difference for the two interfaces is significantly greater than the factor of $\approx$1.4 determined from dc-tuned ST-FMR (Sec.~\ref{subsec:DLT}) and electrically detected spin pumping (Sec.~\ref{subsec:ISH}).
This discrepancy implies that the magnitude of Re$[G_{\uparrow\downarrow}^{eff}]$ of NiFe/Pt calculated from enhanced damping is higher than that calculated for spin injection.  

In addition to spin pumping, interfacial scattering effects~\cite{Rojas-Sanchez2014, Nguyen2014, Park2000, Liu2014c}, e.g., due to proximity-induced magnetization in Pt~\cite{Sun2013a, Ryu2014, Lim2013} or spin-orbit phenomena at the NiFe/Pt interface~\cite{Nembach2014}, may contribute to both stronger damping and lower spin injection in NiFe/Pt.  
Assuming that this interfacial scattering is suppressed by the Cu dusting layer, $\approx$0.010 of $\alpha$ in NiFe/Pt is not accounted for by spin pumping.
The corrected Re$[G_{\uparrow\downarrow}^{eff}]$ for NiFe/Pt is $(8.1\pm1.2)\times10^{14}$~$\Omega^{-1}$m$^{-2}$, which is in excellent agreement with Re$[G_{\uparrow\downarrow}^{eff}]$ calculated from first principles~\cite{Liu2014c}.

Using $G_{\uparrow\downarrow}^{eff}$ quantified above and assuming $\lambda_{Pt}\approx1$ nm~\cite{Boone2013, Boone2014, Kondou2012a, Ganguly2014, Kasai2014, Pai2014a, Obstbaum2014, Skinner2014, Wang2014c}, the intrinsic spin Hall angle $\theta_{SH}$ of Pt and the spin-current transmissivity $T=\theta_{DL}/\theta_{SH}$ across the FM/NM interface can be estimated.  
We obtain $\theta_{SH}\approx0.15$, and $T \approx 0.6$ for NiFe/Pt and $T \approx 0.4$ for NiFe/Cu/Pt. 
These results, in line with a recent report~\cite{Pai2014a}, indicate that the damping-like torque (proportional to $\theta_{DL}$) may be increased by engineering the FM/NM interface, i.e., by increasing $G_{\uparrow\downarrow}^{eff}$.  
For practical applications, the threshold charge current density required for switching or self-oscillation of the magnetization is proportional to the ratio $\alpha/\theta_{DL}$. 
Because of the reciprocity of the damping-like torque and spin pumping, increasing $G_{\uparrow\downarrow}^{eff}$ would also increase $\alpha$ such that it would cancel the benefit of enhancing $\theta_{DL}$.    
Nevertheless, although spin pumping inevitably increases damping, optimal interfacial engineering might minimize damping from interfacial spin-current scattering while maintaining efficient spin-current transmission across the FM/NM interface.  

\subsection{Field-like torque}\label{subsec:FLT}
\begin{figure}
\includegraphics[width=0.7\columnwidth]{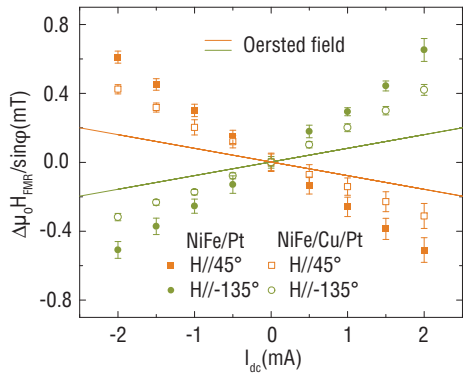}
\caption{\label{fig:fieldlike}Net current-induced effective field, derived from resonance field shift  $\Delta H_{FMR}$ normalized by the field direction angle $|\sin\phi| = 1/\sqrt{2}$. The solid lines denote the estimated Oersted field.} 
\end{figure}

We now quantify the field-like torque from the dc-induced shift in the resonance field $H_{FMR}$, derived from the fit to Eq. \ref{eq:kittel}, as shown in Figs.~\ref{fig:broadband}(b),(c). 
$M_{eff}$ is fixed at its zero-current value so that $\Delta H_{FMR}$ is the only free parameter~\footnote{When $M_{eff}$ is adjustable $M_{eff}$ changes only by $\ll$1\%.}. 
Fig.~\ref{fig:fieldlike} shows the net current-induced effective field, which is equivalent to $\sqrt{2}\Delta H_{FMR}$ in our experimental geometry with the external field applied 45$^\circ$ from the current axis.
The solid lines show the expected Oersted field $\mu_0 H_{Oe} \approx 0.08$ mT per mA for both NiFe/Pt and NiFe/Cu/Pt based on the estimated charge current densities in the NM layers, $H_{Oe} =  \tfrac{1}{2}(J_{c,Pt}t_{Pt}+J_{c,Cu}t_{Cu}-J_{c,Ta}t_{Ta})$, where the contribution from the Pt layer dominates by a factor of $>$6.  

\begin{figure*}
\includegraphics[width=1.2\columnwidth]{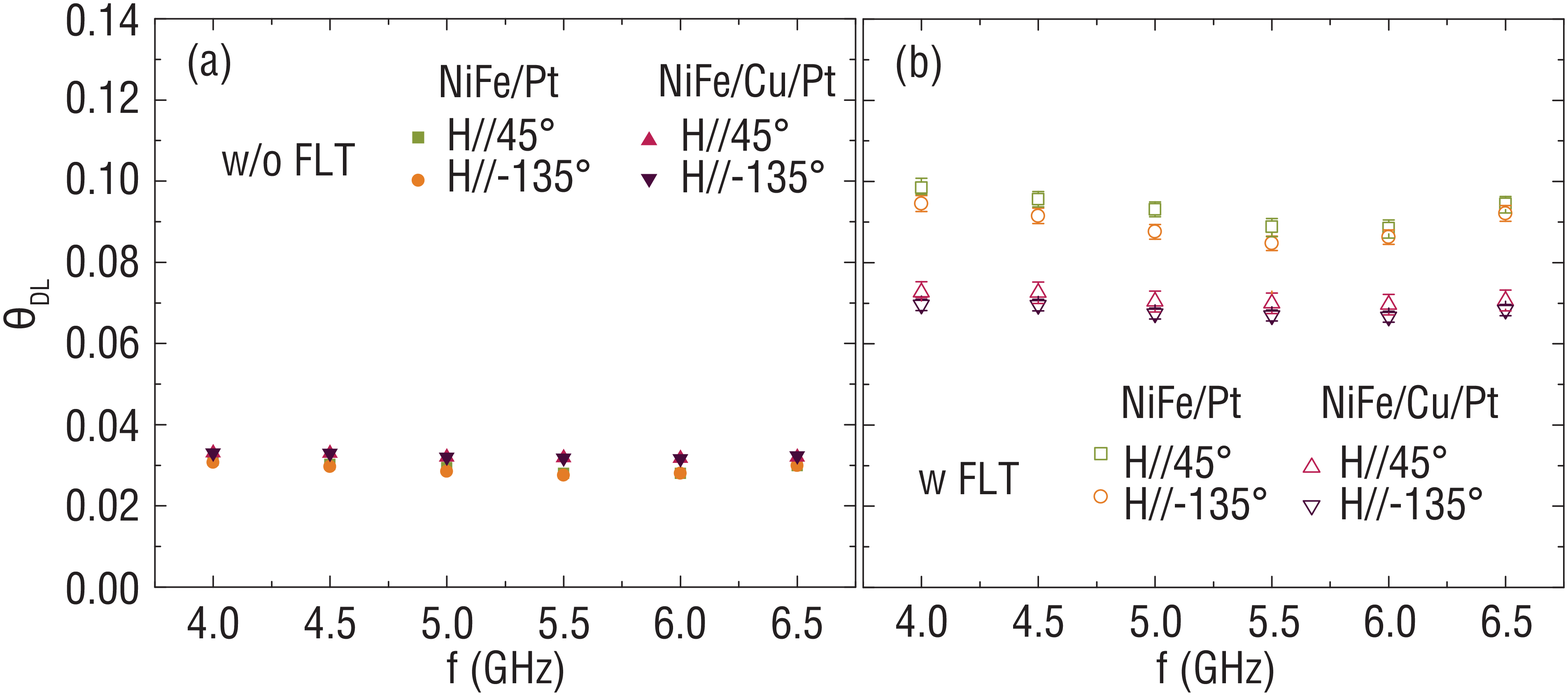}
\caption{\label{fig:lineshape}(a,b) Effective spin Hall angle $\theta^{eff}_{SH,rf}$ extracted from ST-FMR lineshape analysis, disregarding the field-like torque (a) and taking into account the field-like torque (b).} 
\end{figure*}

While the polarity of the shift in $H_{FMR}$ is consistent with the direction of $H_{Oe}$, the magnitude of $\sqrt{2}\Delta H_{FMR}$ exceeds $H_{Oe}$ for both samples as shown in Fig.~\ref{fig:fieldlike}.
This indicates the presence of an additional current-induced effective field due to a field-like torque, $\mu_0 H_{FL} = 0.20\pm0.02$ mT per mA for NiFe/Pt and $\mu_0 H_{FL} = 0.10\pm0.02$ mT per mA for NiFe/Cu/Pt. 
Analogous to $\theta_{DL}$ for the damping-like torque, the field-like torque can also be parameterized by an effective spin Hall angle~\cite{Pai2014a}: 
\begin{equation}\label{eq:thetaFL}
|\theta_{FL}| = \frac{2|e|\mu_0 M_s t_{F}}{\hbar} \left|\frac{H_{FL}}{J_{c,Pt}}\right|.
\end{equation}
Eq.~\ref{eq:thetaFL} yields $\theta_{FL} = 0.024\pm0.003$ for NiFe/Pt and $0.013\pm0.003$ for NiFe/Cu/Pt, comparable to recently reported results in Ref.~\citen{Fan2013}.  

The ultrathin Cu layer at the NiFe/Pt interface reduces the field-like torque by a factor of $1.8\pm0.5$, which is in agreement within experimental uncertainty to the reduction of the damping-like torque (Sec.~\ref{subsec:DLT}).    
This suggests that both torques predominantly originate from the spin Hall effect in Pt.  
Recent studies on FM/NM bilayers using low-frequency measurement techniques~\cite{Fan2013, Fan2014, Pai2014} also suggest that the spin Hall effect is the dominant source of the field-like torque.
Since the field-like torque scales as the imaginary component of $G_{\uparrow\downarrow}^{eff}$ (Refs.~\citen{Haney2013a, Ralph2008, Brataas2012a}), the Cu dusting layer must modify Re[$G_{\uparrow\downarrow}^{eff}$] and Im[$G_{\uparrow\downarrow}^{eff}$] identically.
We estimate $\text{Im}[G_{\uparrow\downarrow}^{eff}] = (\theta_{FL}/\theta_{DL})\text{Re}[G_{\uparrow\downarrow}^{eff}]$ to be $(2.2\pm0.5)\times10^{14}$~$\Omega^{-1}$m$^{-2}$ for NiFe/Pt and $(1.2\pm0.3)\times10^{14}$~$\Omega^{-1}$m$^{-2}$ for NiFe/Cu/Pt. 

Because of the relatively large error bar for the ratio of the field-like torque in NiFe/Pt and NiFe/Cu/Pt, our experimental results do not rule out the existence of another mechanism at the FM/NM interface, distinct from the spin Hall effect.    
For example, the Cu dusting layer may modify the interfacial Rashba effect that can be an additional contribution to the field-like torque~\cite{Gambardella2011a, Haney2013a, Fan2014}.
Also, the upper bound of the field-like torque ratio is close to the factor of $\approx$2 reduction in damping with Cu insertion, possibly suggesting a correlation between the spin-orbit field-like torque and the enhancement in damping at the FM-NM interface.  
Elucidating the exact roles of interfacial spin-orbit effects in FM/HM requires further theoretical and experimental studies.

\subsection{Comparison of the dc-tuned and lineshape methods of ST-FMR}
Accounting for the field-like torque, we determine the effective spin Hall angle $\theta^{rf}_{DL}$ in NiFe/Pt and NiFe/Cu/Pt from the lineshape (Eq.~\ref{eq:lorentzian}) of the ST-FMR spectra at $I_{dc}=0$ (Refs.~\citen{Liu2011, Kondou2012a, Skinner2014, Wang2014c, Mellnik2014, Pai2014a}).  
The coefficients in Eq.~\ref{eq:lorentzian} are $S=V_o\hbar J_{s,rf}/2|e|\mu_0M_st_F$ and $A= V_oH_{rf}\sqrt{1+M_{eff}/H_{FMR}}$, where $V_o$ is the ST-FMR voltage prefactor~\cite{Liu2011} and $H_{rf}\approx\beta J_{c,rf}$ is the net effective rf  magnetic field generated by the rf driving current density $J_{c,rf}$ in the Pt layer. 
$\theta^{rf}_{DL} ={J_{s,rf}}/{J_{c,rf}} $ is calculated from the lineshape coefficients $S$ and $A$:
\begin{equation}\label{eq:SA}
|\theta^{rf}_{DL}| =\left|\frac{S}{A}\right|\frac{2|e|\mu_0M_st_F}{\hbar}\beta \sqrt{1+\frac{M_{eff}}{H_{FMR}}}.
\end{equation}
Fig.~\ref{fig:lineshape}(a) shows $|\theta^{rf}_{DL}|$ obtained by ignoring the field-like torque contribution, i.e., $\beta = t_{Pt}/2$.  
This underestimates $|\theta^{rf}_{DL}|$, implying identical damping-like torques in NiFe/Pt and NiFe/Cu/Pt.  
Using $\beta = t_{Pt}/2+H_{FL}/J_{c,Pt}$ extracted from Fig.~\ref{fig:fieldlike}, $\theta^{rf}_{DL}=0.091\pm0.007$ for NiFe/Pt and $0.069\pm0.005$ for NiFe/Cu/Pt plotted in Fig.~\ref{fig:lineshape}(b) are in agreement with $\theta_{DL}$ determined from the dc-tuned ST-FMR method.  
The presence of a nonnegligible field-like torque in thin FM may account for the underestimation of $\theta^{rf}_{DL}$ based on the lineshape analysis compared to $\theta_{DL}$ from dc-tuned ST-FMR as reported in Refs.~\citen{Ganguly2014, Kasai2014}.

\begin{center}
\begin{table}
\caption{Parameters related to spin-orbit torques} %title of the table
\centering % centering table
\begin{tabular}{c ccccc} % creating eight columns
\hline\hline %inserting double-line
{}&\multicolumn{2}{c}{NiFe/Pt}&\multicolumn{2}{c}{NiFe/Cu/Pt} \\ [0.5ex] 
\hline % inserts single-line
$\theta_{DL}$ & \multicolumn{2}{c}{$0.087\pm0.007$} & \multicolumn{2}{c}{$0.062\pm0.005$}\\
$\theta_{FL}$ & \multicolumn{2}{c}{$0.024\pm0.003$} & \multicolumn{2}{c}{$0.013\pm0.003$}\\
$\text{Re}[G_{\uparrow\downarrow}^{eff}]$ ($10^{14}$ $\Omega^{-1}$m$^{-2}$) & \multicolumn{2}{c}{$8.1\pm1.2$} & \multicolumn{2}{c}{$5.8\pm0.5$}\\
$\text{Im}[G_{\uparrow\downarrow}^{eff}]$ ($10^{14}$ $\Omega^{-1}$m$^{-2}$) & \multicolumn{2}{c}{$2.2\pm0.5$} & \multicolumn{2}{c}{$1.2\pm0.3$}\\
$C_{ISH}\text{Re}[G_{\uparrow\downarrow}^{eff}]$ (a.u.) & \multicolumn{2}{c}{$1.4\pm0.2$} & \multicolumn{2}{c}{$1$}\\
$\alpha-\alpha_0$ & \multicolumn{2}{c}{$0.032\pm0.001$} & \multicolumn{2}{c}{$0.016\pm0.001$}\\[1ex] % [1ex] adds vertical space
\hline % inserts single-line
\end{tabular}
\label{tab:params}
\end{table}
\end{center}
\section{Conclusions}
We have experimentally demonstrated that the spin-orbit damping-like and field-like torques scale with interfacial spin-current transmission.  
Insertion of an ultrathin Cu layer at the NiFe/Pt interface equally reduces the spin-Hall-mediated spin-orbit torques and spin pumping, consistent with diffusive transport of spin current across the FM/NM interface.  
Parameters relevant to spin-orbit torques in NiFe/Pt and NiFe/Cu/Pt quantified in this work are summarized in Table~\ref{tab:params}.
We have also found an additional contribution to damping at the NiFe/Pt interface distinct from spin pumping. 
The dc-tuned ST-FMR technique used here permits precise quantification of spin-orbit torques directly applicable to engineering efficient spin-current-driven devices.  

\begin{figure*}
\includegraphics[width=1.3\columnwidth]{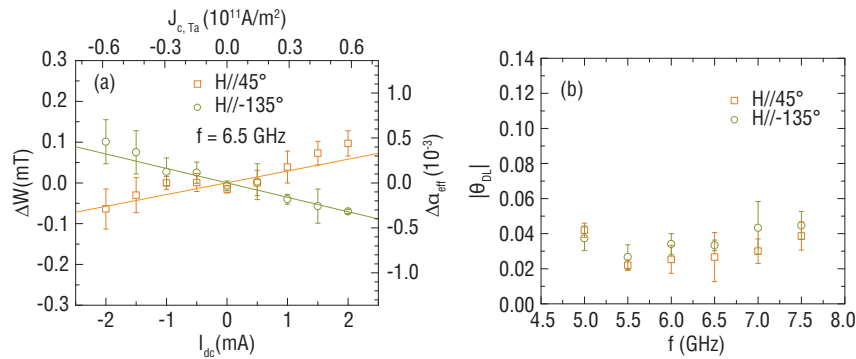}
\caption{\label{fig:Ta}(a) Change in resonance linewidth $W$ versus dc bias current $I_{dc}$ in Ta/NiFe at $f=6.5$ GHz. (b) Effective spin Hall angle $\theta_{DL}$ calculated at several frequencies.} 
\end{figure*}

\section*{Acknowledgements}
T.N. and S.E. contributed equally to this work. 
This work was supported by the Air Force Research Laboratory through contract FA8650-14-C-5706 and in part by FA8650-14-C-5705, the W.M. Keck Foundation, and the National Natural Science Foundation of China (NSFC) 51328203.
Lithography was performed in the George J. Kostas Nanoscale Technology and Manufacturing Research Center.
S.E. thanks Xin Fan and Chi-Feng Pai for helpful discussions.  
T.N. and S.E. thank James Zhou and Brian Chen for assistance in setting up the ST-FMR system, and Vivian Sun for assistance in graphic design. 

\begin{figure*}
\includegraphics[width=0.65\columnwidth]{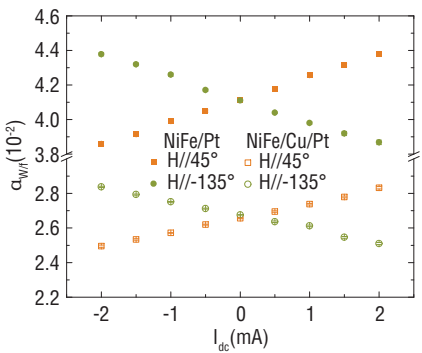}
\caption{\label{fig:alphaIdc}Empirical damping parameter $\alpha_{W/f}$ as a function of dc bias current $I_{dc}$.} 
\end{figure*}
\section*{Appendix A: Damping-like torque contribution from Tantalum}\label{ApA}
With the same dc-tuned ST-FMR technique described in Sec.~\ref{subsec:STFMR}, we evaluate the effective spin Hall angle $\theta_{DL}$ of Ta interfaced with NiFe.  
Because of the high resistivity of Ta, the signal-to-noise ratio of the ST-FMR spectrum is significantly lower than in the case of NiFe/Pt, thus making precise determination of $\theta_{DL}$ more challenging.  
Nevertheless, we are able to obtain an estimate of $\theta_{DL}$ from a 2-$\mu$m wide, 10-$\mu$m long strip of subs/Ta(6 nm)/Ni$_{80}$Fe$_{20}$(4 nm)/Al$_2$O$_3$(1.5 nm) (``Ta/NiFe'') .  
The estimated resistivity of Ta(6 nm) is 200 $\mu\Omega$cm and that of NiFe(4 nm) is 70 $\mu\Omega$cm. 

Fig.~\ref{fig:Ta}(a) shows the change in linewidth $\Delta W$ (or $\Delta \alpha_{eff}$) due to dc bias current $I_{dc}$.  
The polarity of $\Delta W$ against $I_{dc}$ is the same as in NiFe capped with Pt (Fig.~\ref{fig:dampinglike}(a)).  
Because the Ta layer is beneath the NiFe layer, this observed polarity is consistent with the opposite signs of the spin Hall angles for Pt and Ta.  
Here we define the sign of $\theta_{DL}$ for Ta/NiFe to be negative.  
Using Eq.~\ref{eq:JsJc} with $M_s = M_{eff} = 7.0\times10^5$ A/m and averaging the values plotted in Fig.~\ref{fig:Ta}(b), we arrive at $\theta_{DL} = -0.034\pm0.008$.  
This magnitude of $\theta_{DL}$ is substantially smaller than $\theta_{DL} \approx -0.1$ in Ta/CoFe(B)~\cite{Liu2012, Emori2014d} and Ta/FeGaB~\cite{Emori2015}, but similar to reported values of $\theta_{DL}$ in Ta/NiFe bilayers~\cite{Weiler2014c, Deorani2013}.
For the analysis of the damping-like torque in Sec.~\ref{subsec:DLT}, we take into account the $\theta_{DL}$ obtained above and the small charge current density in Ta.  
In the Ta/NiFe/(Cu/)/Pt stacks, owing to the much higher conductivity of Pt, the spin-Hall damping-like torque from the top Pt(4) layer is an order of magnitude greater than the torque from the bottom Ta(3) seed layer. 

\section*{Appendix B: dc dependence of the empirical damping parameter}\label{ApB}
Magnetization dynamics in the presence of an effective field $\mathbf{H}_{eff}$ and a damping-like spin torque is given by the Landau-Lifshitz-Gilbert-Slonczewski equation:
\begin{equation}
\frac{\partial \mathbf{m}}{\partial t} = -|\gamma|{\mathbf{m}}\times\mathbf{H}_{eff}
+\alpha\mathbf{m}\times\frac{\partial \mathbf{m}}{\partial t}
+\tau_{DL}\mathbf{m}\times(\boldsymbol{\sigma}\times\mathbf{m}),
\end{equation}
where $\tau_{DL}$ is a coefficient for the damping-like torque (proportional to $\theta_{DL}$) and $\boldsymbol{\sigma}$ is the orientation of the spin moment entering the FM. 
%+\frac{|\gamma|}{M_s}\vec\nabla \cdot \mathbf{J}_s.
Within this theoretical framework, it is not possible to come up with a single Gilbert damping parameter as a function of bias dc current $I_{dc}$ that holds at all frequencies.  
However, at $I_{dc} = 0$ we empirically extract the damping parameter $\alpha$ from the linear relationship of linewidth $W$ versus frequency $f$ (Eq.~\ref{eq:damping}).
We can take the same approach and define an empirical damping parameter $\alpha_{W/f}$ as a function of $I_{dc}$, i.e. 
\begin{equation}\label{eq:dampingIdc}
W(I_{dc}) = W_0 + \frac{2\pi\alpha_{W/f}(I_{dc})}{|\gamma|}f,
\end{equation}
where we fix the inhomogeneous linewidth broadening $W_0$ at the value at $I_{dc} = 0$, which does not change systematically as a function of small $I_{dc}$ used here.  
This approach of setting $\alpha_{W/f}$ as the only fitting parameter in Eq.~\ref{eq:dampingIdc} well describes our data (e.g., Fig.~\ref{fig:broadband}(a)).  
We show in Fig.~\ref{fig:alphaIdc} the resulting $\alpha_{W/f}$ versus $I_{dc}$.  
The change in $\alpha_{W/f}$ normalized by the charge current density in Pt is $0.0036\pm0.0001$ per $10^{11}$ A/m$^2$ for NiFe/Pt and $0.0025\pm0.0001$ per $10^{11}$ A/m$^2$ for NiFe/Cu/Pt.
This empirical measure of the damping-like torque again exhibits a factor of $\approx$1.4 difference between NiFe/Pt and NiFe/Cu/Pt.

\end{document}